\documentclass[twocolumn,twoside]{revtex4}
\usepackage{siunitx}
\usepackage{graphicx}
\usepackage{fancyhdr}
\usepackage{amsmath}
\usepackage{tikz}
\begin{document}

\title{The Search for Electric Dipole Moments of Charged Particles in Storage Rings}

%

\author{A.~Andres\\
on behalf of the JEDI collaboration}
\affiliation{Institut f\"ur Kernphysik, Forschungszentrum J\"ulich, 52425 J\"ulich, Germany \\
III. Physikalisches Institut B, RWTH Aachen University, 52056 Aachen, Germany
} 

\begin{abstract}
The matter-antimatter asymmetry cannot be explained by the Standard Model (SM) of elementary particle physics. According to A. Sakharov, additional sources of $\mathcal{CP}$-Violating phenomena are needed to understand the matter-antimatter asymmetry. Electric Dipole Moments (EDMs) of subatomic elementary particles may provide additional $\mathcal{CP}$ violation, since they violate $\mathcal{T}$ (and $\mathcal{P}$) symmetry. Polarized beams in storage rings offer the possibility to measure EDMs of charged particles by observing the influence of the EDM on the spin motion. The Cooler Synchrotron (COSY) at Forschungszentrum J\"ulich provides polarized protons and deuterons up to a momentum of \SI{3.7}{GeV/c} and is therefore an ideal starting point for the JEDI - Collaboration (J\"ulich Electric Dipole moment Investigations) to perform the first direct measurement of the deuteron EDM. This document describes recent results of EDM activities at COSY.
\end{abstract}

\maketitle

\thispagestyle{fancy}


\section{Introduction}
\begin{figure}[tbh]
\centering
\includegraphics[width=80mm]{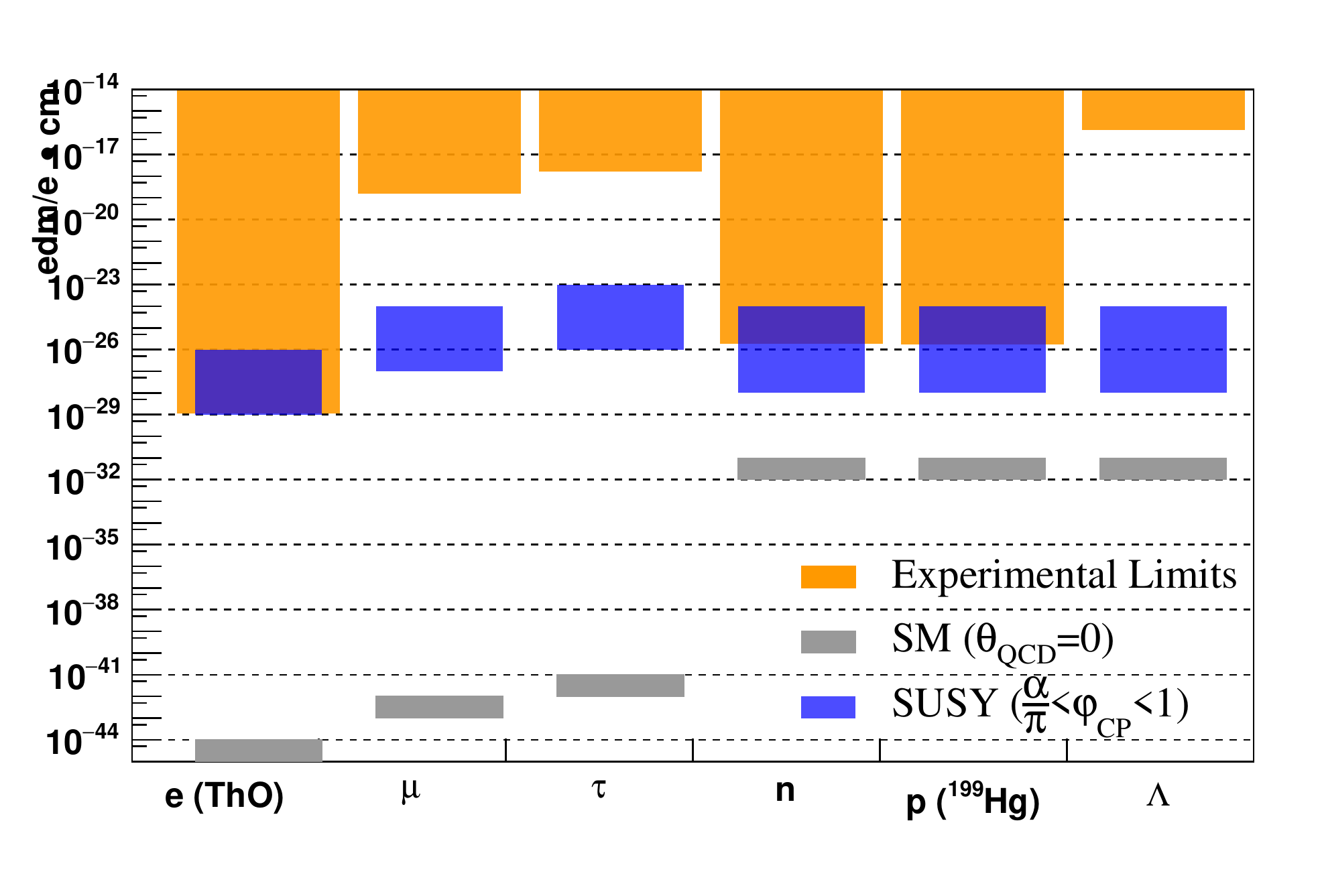}
\caption{90\% EDM limits for various particles, together with predictions from the Standard Model and Super Symmetry.} \label{fig:current_edm_limits}
\end{figure}

The observable matter-antimatter asymmetry in our post-Big-Bang universe cannot be explained by the Standard Model of elementary particle physics \cite{edm_cp}. Additional sources of $\mathcal{CP}$ violation could explain the excess of matter in the universe \cite{edm_cp}. Electric Dipole Moments of subatomic elementary particles can only exist if parity ($\mathcal{P}$) and time reversal ($\mathcal{T}$) symmetry are violated. If the $\mathcal{CPT}$ theorem holds, $\mathcal{T}$-violation implies $\mathcal{CP}$ violation. The EDM is a vectorial property of a elementary particle, aligned with its spin. The predictions of EDMs of the Standard Model are orders of magnitude too small to explain the dominance of matter in the universe. A discovery of a larger EDM would hint towards physics beyond the SM and contribute to an explanation for the dominance of matter in the universe \cite{wirzba}. EDM searches have already been conducted on neutron confined traps. Furthermore, EDM limits for electrons, muons and tauons have been obtained directly. However, for the proton and deuteron, no direct measurements are available up to date. The JEDI collaboration is planning to accomplish a first direct measurement of the deuteron EDM at the COSY accelerator facility. In Figure \ref{fig:current_edm_limits}, current EDM limits for different particles are summarized \cite{wirzba,g-2}. All measurements are consistent with 0. 

In chapter \ref{chap:COSY} the basics concepts of measuring an EDM in storage rings is introduced while in chapter \ref{chap:permanent_edm} and \ref{chap:oscillating_edm} the measurements of the permanent deuteron EDM and the oscillating axion/ axion like particles (ALPs) are discussed. Chapter \ref{chap:future} deals with future EDM activities in storage rings foreseen by the JEDI-collaboration. 

\section{Spin Motion in a Storage Ring}
\label{chap:COSY}
The evolution of the spin vector (i.e beam polarization vector) in an electromagnetic field, including a non vanishing electric dipole moment, is described by the Thomas-BMT equation. Assuming that the particles motion $\vec{v}$ is perpendicular to the electric and magnetic field, the T-BMT equation is given by \cite{TBMT}
\begin{align}
\frac{\mathrm{d}\vec{s}}{\mathrm{d}t} &= \left(\vec{\Omega}^{\mathrm{MDM}}+\vec{\Omega}^{\mathrm{EDM}}\right) \times \vec{s}  \;\;\; \text{with} \label{eq:TBMT}\\
\vec{\Omega}^{\mathrm{MDM}} &=-\frac{q}{m}\left[G \vec{B}+\left(G-\frac{1}{\gamma^{2}-1}\right) c \vec{\beta} \times \vec{E} \right], \label{eq:MDM} \\
\vec{\Omega}^{\mathrm{EDM}} &=-\frac{q}{m}\left[\frac{\eta}{2} \left(\vec{E}+c\vec{\beta} \times \vec{B}\right)\right]. \label{eq:EDM}
\end{align}
In Eq. (\ref{eq:TBMT}), $\vec{\Omega}^{\mathrm{MDM}}$ and $\vec{\Omega}^{\mathrm{EDM}}$ denote the angular velocities associated with the magnetic (MDM) and the electric dipole moment (EDM). $\vec{E}$ and $\vec{B}$ are external electric and magnetic fields in the laboratory frame, \mbox{$G=(g-2)/2$} is the anomalous $g$-factor and $\eta$ is a dimensionless parameter describing the strength of the EDM. The Lorentz factors $\gamma$ and $\vec{\beta}$ are used to describe the relativistic velocity of particles. $\vec{\beta}$ is defined as the ratio of the particles velocity $\vec{v}$ and the speed of light in vacuum $\vec{\beta} = \vec{v}/c_0$. Using $\vec{\beta}$, $\gamma$ can be calculated via $\gamma = 1/\sqrt{1-|\vec{\beta}|^2}$. Relevant numbers for the experiment are summarized in Table \ref{tab:experiment_parameters}.
According to Eq. (\ref{eq:EDM}), in the presence of an EDM, the vertical polarization component oscillates with a small amplitude $\beta\eta/(2G)$ and an angular frequency $|\vec{\Omega}^{\text{MDM}}| = -qGB/m$. This oscillation is used by the $g-2$ experiment to measure a limit for the muon EDM \cite{g-2}. However, as the $G$ factors for leptons are approximately $10^{-3}$, while the $G$ factor for hadrons is $\mathcal{O}(1)$, other methods need to be developed for hadron EDM experiments. 
To avoid cancellation of EDM effects, a new radio-frequency (RF) Wien filter was designed, built and commissioned in the ring which operates on a resonant frequency of the spin precession $f_{\text{WF}} \approx \SI{873}{kHz}$. A Wien filter provides a horizontal electric field and a vertical magnetic field such that the Lorentz force on the beam vanishes and the magnetic and electric field only act on the polarization vector. It converts the microscopic oscillation of the polarization into a macroscopic vertical build up \cite{wien_filter}.

\begin{table}[]
\begin{center}
\caption{Relevant parameters for the deuteron EDM experiment at COSY. The $G$ factor can be found in reference \cite{g_factor}.}
\begin{tabular}{lcl}
\hline
\hline
\multicolumn{1}{l|}{deuteron momentum }   &\multicolumn{1}{l|}{$p$}& \SI{0.970}{GeV/c}   \\
\multicolumn{1}{l|}{rel. velocity}        &\multicolumn{1}{l|}{$\beta$}& 0.459           \\
\multicolumn{1}{l|}{Lorentz Factor}       &\multicolumn{1}{l|}{ $\gamma$}& 1.126         \\
\multicolumn{1}{l|}{gyromagnetic anomaly} &\multicolumn{1}{l|}{$G$}& $\approx -0.143$    \\
\multicolumn{1}{l|}{revolution frequency} &\multicolumn{1}{l|}{ $f_{\text{cosy}}$}& \SI{752543}{Hz} \\
\multicolumn{1}{l|}{precession frequency} &\multicolumn{1}{l|}{$|\vec{\Omega}^{\text{MDM}}|$}& \SI{121173}{Hz} \\
\multicolumn{1}{l|}{cycle length}         &\multicolumn{1}{l|}{}& \SI{270}{s}     \\
\multicolumn{1}{l|}{RF Wien Filter}       &\multicolumn{1}{l|}{$f_{\text{WF}}$}& \SI{872949}{Hz} \\ \hline
\hline
                                                            &                                     
\end{tabular}
\label{tab:experiment_parameters}
\end{center}
\end{table}

\section{First Deuteron EDM Precursor measurement at COSY}
\label{chap:permanent_edm}
The JEDI collaboration \cite{jedi} has proposed a staged approach to measure the deuteron EDM directly for the first time. The Cooler Synchrotron (COSY) located at Forschungszentrum Jülich is an ideal starting point for these activities. More information about future EDM projects is given in chapter \ref{chap:future}.
\subsection{The Cooler Synchrotron COSY}
COSY is a race-track shaped storage ring with a circumference of \SI{184}{m}, which provides phase-space cooled polarized and unpolarized deuteron and proton beams up to a momentum of \SI{3.7}{GeV}. 

The JEDI collaboration performed measurements in November 2018 and March 2021 to measure for the first time directly the deuteron EDM.  

In addition, spring 2019, the JEDI collaboration performed a first proof-of-principle experiment at COSY to search for axions/ axion like particles and an oscillating EDM.

Both experiments require a long in-plane polarization time (also called Spin Coherence Time). At COSY $1/e$ spin coherence times over \SI{1000}{s} have been achieved for deuterons \cite{SCT}.

\subsection{Methodology}
\begin{figure}[h]
\centering
\includegraphics[width=80mm]{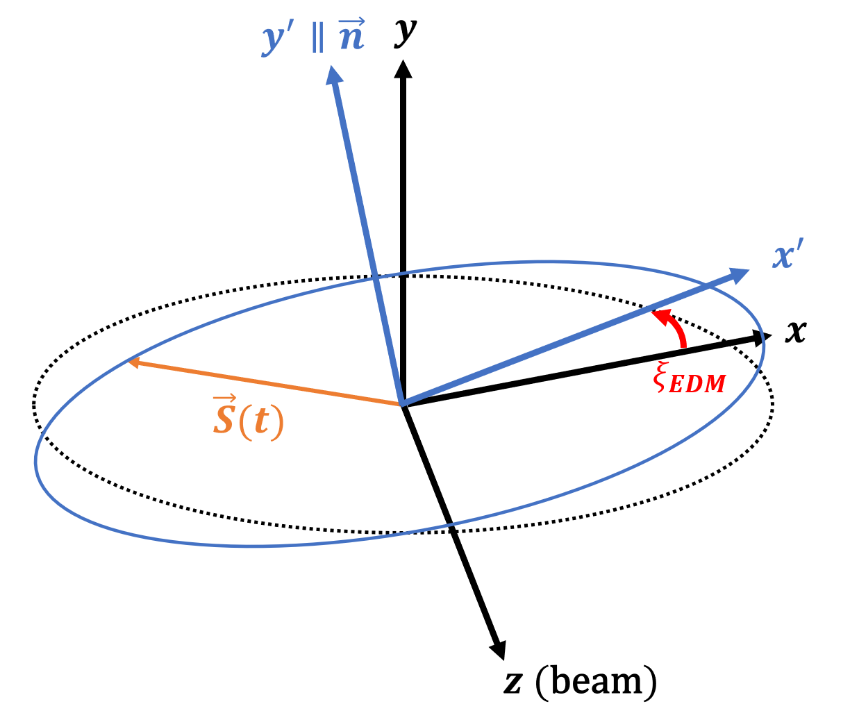}
\caption{Coordinate system: $z$ denotes the longitudinal and beam direction. $x$ points in radial direction and $y$ points perpendicular to the accelerator plane ($x-z$). The EDM tilts the invariant spin axis $y$ in radial direction by $\xi_{\text{EDM}}$.} \label{fig:rotation_zeta}
\end{figure}

The main idea of the experiment is to measure the influence of the EDM on the beam polarization. A fully vertically polarized deuteron beam is injected into the ring, accelerated to a momentum \SI{970}{MeV}, bunched and cooled. The beam polarization is measured with a polarimeter by scattering a fraction of the particles on a carbon target \cite{jepo}. After the polarization is rotated into the horizontal plane (= accelerator plane), the polarization starts to rotate around the so-called invariant spin axis with \mbox{$|\vec{\Omega}^{\text{MDM}}| = |\gamma G|(\cdot f_{\text{cosy}} \approx \SI{121}{kHz}$}) ($y$ axis in Figure \ref{fig:rotation_zeta}) due to the vertical $\vec{B}$ fields in the dipole magnets. The number of rotations of the polarization per turn in the storage ring is called spin tune and given by $\nu_s = \gamma G \approx -0.16$. The negative sign stands for a counter-clockwise rotation of the polarization while the particles are rotating clockwise in the storage ring. In an ideal accelerator without an EDM, the invariant spin axis would be parallel to the vertical magnetic dipole fields. The existence of an EDM tilts the invariant spin axis in radial direction ($x$ direction in \mbox{Figure \ref{fig:rotation_zeta})} by an angle $\xi_{\text{EDM}}$. The EDM strength $\eta$ can be calculated via
\begin{equation}
	\xi_{\text{EDM}} = \arctan\left(\frac{\eta\beta}{2G}\right).
\end{equation}

\begin{figure}[h]
\centering
\begin{tikzpicture}
    \draw(0,0) node{\includegraphics[width=80mm]{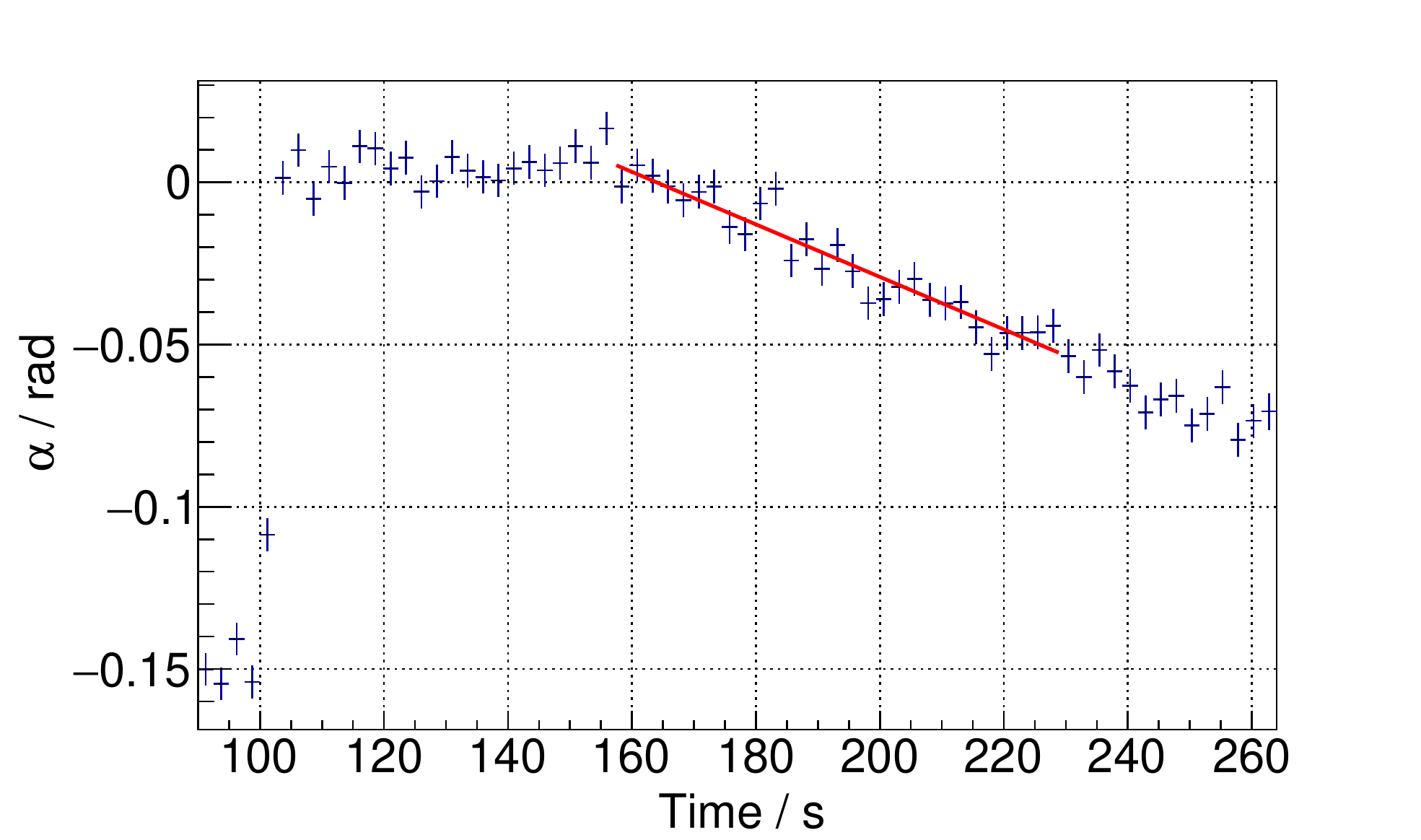}};
    \draw(1.4,-1.5) node{Wien filter \& Snake On};
    \draw (-0.55,-1.75) -- (-0.55,1.9);
    \end{tikzpicture}
\caption{Angle between vertical and horizontal polarization component $\alpha = \arctan\left(P_v/P_h\right)$ as a function of time. The vertically polarized beam is rotated into the horizontal plane at $t = \SI{100}{s}$. When Snake and RF Wien filter are switched on at $t = \SI{155}{s}$, the vertical polarization accumulates linearly. This build up was measured at the following settings: $\phi^{\text{WF}} = \SI{0.945}{mrad}$, $\xi^{\text{SOL}} = \SI{0}{mrad}$ and $\phi_{\text{rel}} = \SI{0.79}{rad}$.} \label{fig:build_up}
\end{figure}

\begin{figure}[h]
\centering
\includegraphics[width=80mm]{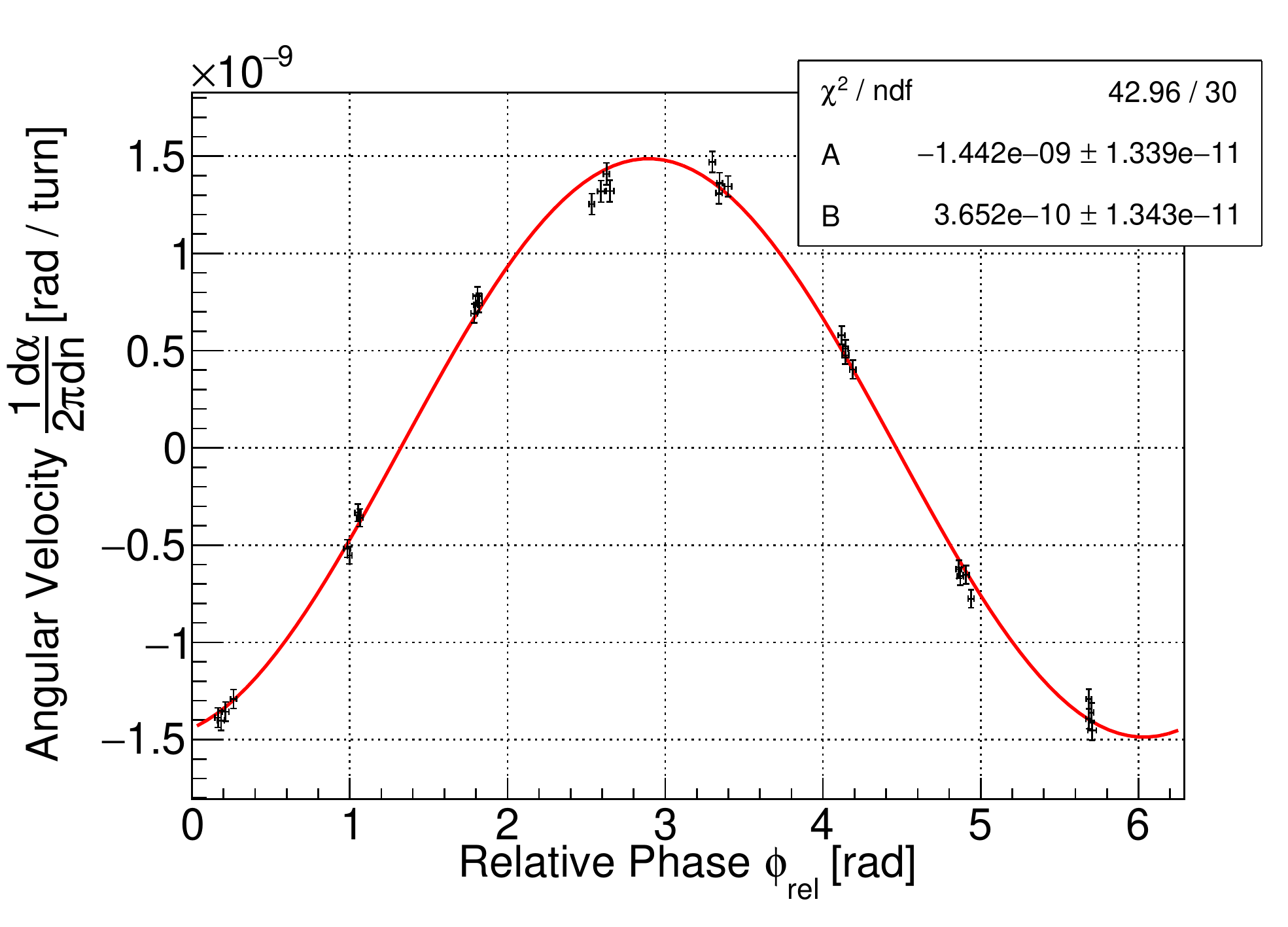}
\caption{Sinusoidal dependence of the linear build up of the vertical polarization as a function of the relative phase $\phi_{\text{rel}}$. The amplitude of the sinusoidal fit function corresponds to the resonance strength $\epsilon$. All measurements are done with $\phi^{\text{WF}} = \SI{0.945}{mrad}$ and $\xi^{\text{SOL}} = \SI{0}{mrad}$. } \label{fig:rel_phase_build_up}
\end{figure}

\begin{figure}[h]
\centering
\includegraphics[width=80mm]{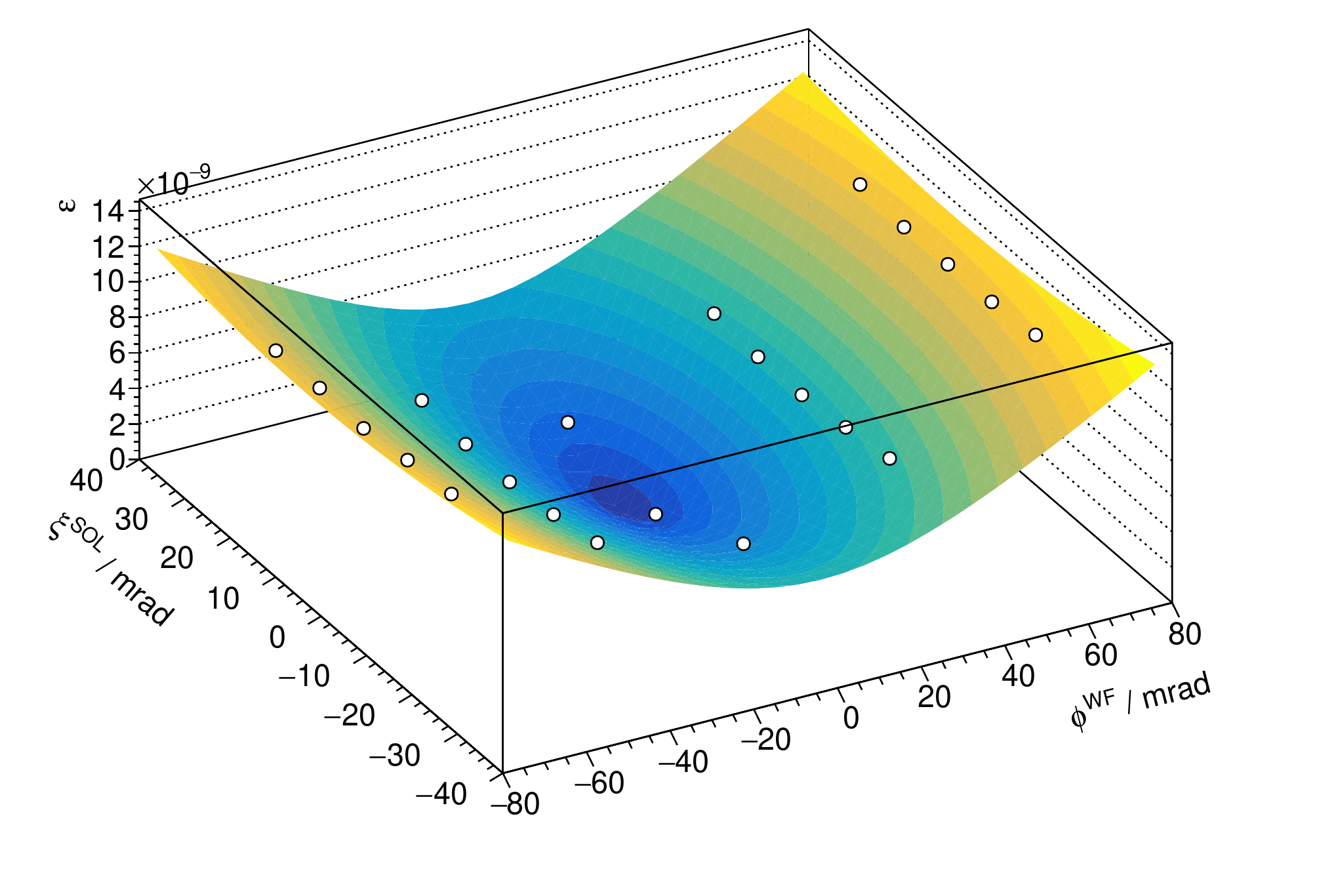}
\caption{Resonance Strengths $\epsilon$ as a function of Wien Filter rotation angle and spin rotation angle inside the Siberian Snake. The minimum of the surface gives the orientation of the invariant spin axis in radial and longitudinal direction.} \label{fig:map}
\end{figure}

The main goal of the experiment is to measure the orientation of the invariant spin axis which gives direct access to the EDM. Two devices are installed in the ring to help to measure the orientation of the invariant spin axis. The RF Wien filter was introduced in chapter \ref{chap:COSY}. The RF Wien filter adds a $x$-component (roll) to the orientation of the invariant spin axis by rotation the RF Wien filter around the beam axis by $\phi_{\text{WF}}$. A so called Siberian Snake provides a solenoidal field magnetic field which is used to tilt the invariant spin axis in longitudinal $z$-direction $\xi^{\text{SOL}}$(pitch).
In order to see a build up of the vertical polarization, the RF Wien Filter needs to run on resonance with the spin precession frequency
\begin{align}
	E_{\text{WF}} &= E_0\cos(2\pi f_{\text{cosy}}|k+\nu_{s}| + \phi_{\text{rel}}), \\
	B_{\text{WF}} &= B_0\cos(2\pi f_{\text{cosy}}|k+\nu_{s}| + \phi_{\text{rel}}). 
\end{align}
A so-called phase-feedback was developed by the JEDI collaboration \cite{phase_feedback} which changes in real-time the RF Wien Filter frequency to keep the spin precession frequency and Wien filter frequency in phase. The Wien filter is changing its fields on one of the harmonics (\mbox{$k=-1$}) of the spin precession frequency $\nu_{s}$ so that the particles passing through the device receive a spin kick in the same direction each turn. The magnitude of the linear build-up of the vertical polarization over time depends on the chosen relative phase $\phi_{\text{rel}}$. \mbox{Figure \ref{fig:build_up}} shows an experimental result for the linear build up of the angle between vertical and horizontal polarization ($\alpha = \arctan(P_v/P_h)$) as a function of time after switching on RF Wien filter and Siberian Snake at $t = \SI{155}{s}$.
\mbox{Figure \ref{fig:rel_phase_build_up}} shows the sinusoidal dependence of the build up as a function of the relative phase $\phi_{\text{rel}}$. The out-of-plane angles are measured for 8 relative phases. The slope ($\text{d}\alpha/\text{d}t$) as shown in \mbox{Figure \ref{fig:build_up}} is scaled by $2\pi f_{\text{cosy}}$ which translates the slope from build up per time unit ($t$) to build up per turn ($n$) in the accelerator ($\text{d}\alpha/\text{d}n$). The sinusoidal dependence of the angular velocity as a function of the relative phase is fitted with
\begin{equation}
	\frac{1}{2\pi}\frac{\text{d}\alpha}{\text{d}n} = A \sin(\phi_{\text{rel}}) + B \cos(\phi_{\text{rel}}).
\end{equation}
The resonance strength $\epsilon$, i.e the maximum build up rate, can be obtained from the fit parameters 
\begin{equation}
	\epsilon = \sqrt{A^2+B^2}.
\end{equation}
In Figure \ref{fig:map}, the resonance strength is shown as a function of Wien filter rotation angle and spin rotation angle inside the Siberian Snake. A theoretical description of the resonance strength is given by \cite{resonance_strength}
\begin{multline}
	\epsilon(\phi^{\text{WF}},\xi^{\text{Sol}})=\bigg[A_{\text{WF}}^{2}\left(\phi^{\text{WF}}-\phi_{0}^{\text{WF}}\right)^{2}\\+\frac{A_{\text{Sol}}^{2}}{4\sin^2{(\pi \nu_{s})}}\left(\xi_{0}^{\text{Sol}}-\xi^{\text{Sol}}\right)^{2}\bigg]^{\frac{1}{2}}.
\label{eq:epsilon_fitfunc}
\end{multline}
The overall shape of Eq. (\ref{eq:epsilon_fitfunc}) is parabolic in radial ($\phi^{\text{WF}}$) and longitudinal ($\xi^{\text{SOL}}$) direction. $A_{\text{WF}}$ and $A_{\text{Sol}}$ are scaling factors for the paraboloids and $\nu_{s}$ is the unperturbed spin tune. The fit parameters $\phi_0^{\text{WF}}$ and $\xi_0^{\text{SOL}}$ denote the minima of the paraboloids in radial and longitudinal direction and the direction of the invariant spin axis in $x$ and $z$ direction in \mbox{Figure \ref{fig:rotation_zeta}}, respectively. The minimum according to a fit to the data points reads
\begin{align*}
    \phi_0^{\text{WF}} & = \SI{-2.91 \pm 0.08}{mrad}\\
    \xi_0^{\text{SOL}} & = \SI{-5.22 \pm 0.07}{mrad}
\end{align*}
In the ideal case of a perfectly aligned storage ring, the invariant spin axis should only be tilted in radial direction ($\phi^{\text{WF}}$) due to an EDM. However, according to the fit, the invariant spin axis is also tilted in longitudinal and radial direction by a few mrad, caused by systematic effects (e.g. misalignments of elements) which are currently under investigation using beam and spin tracking simulations. Note that a tilt in radial direction of \SI{1}{mrad} is equivalent to an EDM of $10^{-17}\;$ecm.
\section{Search for Axions/ Axion like particles (ALPs)}
\label{chap:oscillating_edm}

\begin{figure}[h]
\centering
\includegraphics[width=80mm]{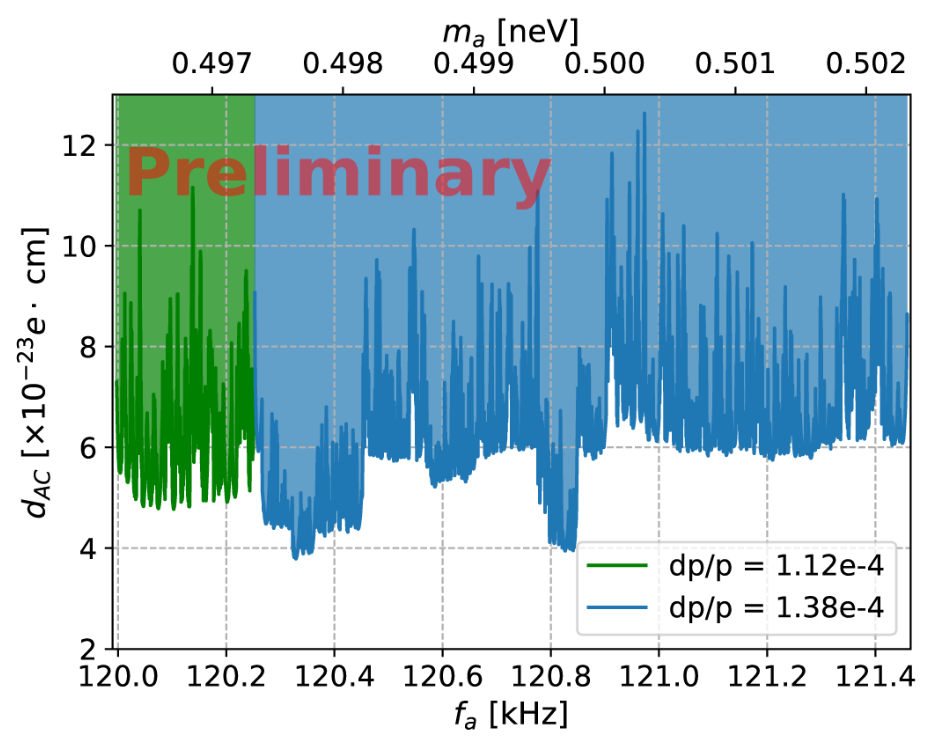}
\caption{90\% EDM confidence limits for the oscillating EDM component ($d_1$) in Eq. \ref{eq:osc_axion_edm} ($d_{AC}$ in Figure). The frequency $f_a$ corresponds to the MDM spin precession frequency $\Omega^{\text{MDM}}$ and can be converted to the axion mass $m_a$ by using Eq. (\ref{eq:osc_axion_edm_calc}). The green and blue areas mark two different ramping rates for the beam momentum (i.e $\gamma$). } \label{fig:current_axion_edm_limits}
\end{figure}

Axions or ALPs were introduced to solve the strong CP problem in quantum chromodynamics \cite{axion_strong_cp}. They are also a candidate for dark matter in the universe. The existence of an axion or an axion like Particle (ALP) would induce an oscillating EDM, where the oscillation frequency is proportional to the axion/ALP mass $m_a$ if it is coupled to gluons \cite{osc_edm_1,osc_edm_2}

\begin{align}
	d(t) &=  d_0 + d_1\sin(\omega_at+\varphi_a)\label{eq:osc_axion_edm} \\ 
	\text{with } \omega_a &= \frac{m_ac^2}{\hbar}, \label{eq:osc_axion_edm_calc}
\end{align}
where the permanent EDM $d_0$ is discussed in section \ref{chap:permanent_edm}. In case of an oscillating EDM \mbox{$d_1\neq0$}, a vertical polarization builds up if the resonance condition
\begin{equation}
	\frac{c^2m_a}{\pi} = \left|\vec{\Omega}^{\text{MDM}}\right| \stackrel{\text{COSY}}{=} \gamma G f_{\text{COSY}}
\end{equation}
is fulfilled. In case of a magnetic ring like COSY, the magnetic dipole moment contribution to the spin precession $\vec{\Omega}^{\text{MDM}}$ is given by $\gamma G$ (see Eq. (\ref{eq:MDM})). Thus, by changing the particle beam energy ($\gamma$), the axion mass can be scanned. In a proof-of-principle experiment at COSY in spring 2019, the JEDI collaboration scanned a mass range equivalent of \SI{4.95}{} to $5.02\cdot10^{-9}\;$eV. Four bunches are simultaneously stored in the machine, each with a $\pi/2\;$rad polarization phase shift with respect to each other. This way the problem of the unknown phase $\varphi_a$ of the axion field with respect to the spin precession phase shall be overcome. In Figure \ref{fig:current_axion_edm_limits}, the 90\% confidence limits calculated from the statistical uncertainty of the vertical polarization is shown. No signal is observed. The sensitivity of the oscillating EDM component is approximately $10^{-23}\;$eV.  

\section{Design of Future Dedicated Storage Rings}
\label{chap:future}
To reduce systematics to the same level as statistics for the search of EDMs, new, dedicated storage rings are needed which have counter and counter-clockwise running beams running in the so-called frozen spin mode $|\vec{\Omega}^{\text{MDM}}| = 0$, i.e the spin precession due to the magnetic moment is suppressed and the beam polarization stays aligned with the beam direction.

The next stage for the EDM measurement in storage ring is to build the so-called Prototype Storage Ring (PSR) using combined $E/B$ fields to bend the beam in a configuration, so that $\Omega^{\text{MDM}} = \vec{0}$. The bending radius for this combined ring is $r\approx\SI{9}{m}$ with field strengths of $|\vec{E}| \approx \SI{7}{MeV/m}$ and $|\vec{B}|\approx\SI{0.03}{T}$. To ensure counter and counter-clockwise operation, the magnetic field has to be reversed. More information about the PSR and the CPEDM (Charged Particle EDM) collaboration can be found in reference \cite{prototype}.

The final goal is to build a ring using only electrostatic benders. In this case, all $B$ field contributions in Equation (\ref{eq:MDM}) vanish. Note that only particles with $G>0$ can be used to fulfil the frozen spin condition in an all-electric ring. The main advantage is the operation of clockwise and counter-clockwise running beams. The main disadvantage is the size of the ring. Assuming electrostatic benders with $E = \SI{8}{MeV/m}$, the bending radius of the ring is $r = \SI{50}{m}$.

\section{summary \& Conclusion}
The search for an EDM aligned with the particles spin axis is of fundamental importance. High precision experiments at accelerators like COSY are ideal to directly measure EDMs of charged particles. At COSY first proof-of-principle measurements for permanent EDMs of deuterons and oscillating EDMs of axions have been performed. As a next step, new dedicated particle accelerators have to be build to increase the systematic accuracy to the same level as statistical uncertainties. This work was supported by the ERC Advanced Grant (srEDM \#694340) of the European Union.


\bigskip 
\begin{thebibliography}{99}

\bibitem{edm_cp} A. Sakharov, Violation of CP invariance, C asymmetry, and baryon asymmetry of the
universe, JETP Lett., vol. 5, pp. 24-27, Jan. 1967

\bibitem{wirzba} Andreas Wirzba, Jan Bsaisou, and Andreas Nogga. Permanent Electric Dipole Moments of Single-, Two- and Three-Nucleon Systems. Int. J. Mod. Phys., E26(01n02):1740031, 2017.

\bibitem{g_factor} H. Masui and M. Kimura, Deuteron-like neutron–proton correlation in 18F studied with the cluster-orbital shell model approach, Progress of The-
oretical and Experimental Physics, 2016 (2016). 053D01.

\bibitem{g-2} G. W. Bennett et al. An Improved Limit on the Muon Electric Dipole Moment. Phys. Rev., D80:052008,
2009

\bibitem{wien_filter} J. Slim, First commissioning results of the waveguide RF Wien filter, Hyperfine Interact 240:7, 2019.

\bibitem{jedi} JEDI collaboration \url{http://collaborations.fz-juelich.de/ikp/jedi}

\bibitem{TBMT} V. Bargmann, Louis Michel, and V. L. Telegdi. Precession of the polarization of particles moving in a
homogeneous electromagnetic field. \textit{Phys. Rev. Lett.}, 2:435–436, 1959

\bibitem{SCT} G. Guidoboni, How to Reach a Thousand-Second in-Plane Polarization Lifetime with 0.97-GeV/c Deuterons in a Storage Ring. Phys. Rev. Lett. 117, 054801, 2016.

\bibitem{phase_feedback} N. Hempelmann et al. Phase Locking the Spin Precession in a Storage Ring. Phys. Rev. Lett. 119, 014801, 2017.

\bibitem{prototype} Abusaif, F. et al. \textit{Storage ring to search for electric dipole moments of charged particles: Feasibility study}.
CERN Yellow Reports: Monographs. CERN, Geneva, 2021.

\bibitem{resonance_strength} F. Rathmann, N. N. Nikolaev and J. Slim, Spin dynamics investigations for the electric
dipole moment experiment, Physical Review Accelerators and Beams 23, 024601, 2020.

\bibitem{axion_strong_cp} R. D. Peccei and H. R. Quinn, “CP conservation in the presence of pseudoparticles”, Phys. Rev. Lett., vol. 38, pp. 1440–
1443, Jun. 1977. doi:10.1103/PhysRevLett.38.1440

\bibitem{osc_edm_1} P. W. Graham and S. Rajendran, “Axion dark matter detection with cold molecules”, Phys. Rev. D, vol. 84, p. 055013, Sep.
2011. doi:10.1103/PhysRevD.84.055013

\bibitem{osc_edm_2} P. W. Graham and S. Rajendran, “New observables for direct detection of axion dark matter”, Phys. Rev. D, vol. 88, p.
035023, Aug. 2013. doi:10.1103/PhysRevD.88.035023

\bibitem{jepo} F. Müller et al, A new beam polarimeter at COSY to search for electric dipole moments of charged particles, JINST 15 P12005, 2020.

\end{thebibliography}
\end{document}